\documentclass{raa}         
\usepackage{graphicx,times}
\usepackage{natbib}
\usepackage{amssymb,amsmath}
\bibpunct{(}{)}{;}{a}{}{,}

\usepackage[a4paper=true,dvipdfm=true,pagebackref=true]{hyperref}
\hypersetup{pdftitle = The title of my PDF, pdfauthor = My name, pdfsubject= The subject, pdfkeywords = keyword1 keyword2} 
\hypersetup{colorlinks = true, linkcolor = green, anchorcolor = red, citecolor = blue, filecolor = red, pagecolor = red, urlcolor = red}

\begin{document}
\title { Variations of the harmonic components of the X-ray Pulse Profile of PSR B1509-58}
\volnopage{Vol.0 (200x) No.0, 000--000}      
   \setcounter{page}{1}          

\author{Pragati Pradhan
      \inst{1,2}
   \and Biswajit Paul
      \inst{3}
   \and Harsha Raichur
      \inst{3}
  \and B.C Paul
  \inst{2}
   }

   \institute{St Joseph's College, Singamari, Darjeeling 734104, West Bengal, India ; {\it pragati2707@gmail.com} \\
        \and
             North Bengal University, Raja Rammohanpur, District Darjeeling 734013, West Bengal, India\\
        \and
             Raman Research Institute, Sadashivnagar, Bangalore 560080, India\\
   }

\abstract
{We used the Fourier decomposition technique to investigate the stability of the X-ray pulse profile
of a young pulsar 
PSR B1509-58 by studying the relative amplitudes and the
phase differences of its harmonic components with respect to the fundamental using data from the Rossi X-Ray Timing Explorer. 
Like most young rotation powered pulsars, PSR B1509-58 has a high spin down rate. 
It also has less timing noise
 allowing accurate measurement of higher order 
frequency derivatives which in turn helps in study of the physics of pulsar spin down. Detailed investigation of pulse profiles
over the years will help us establish any possible connection between the timing characteristics and the high energy emission 
characteristics for this pulsar. Further, the 
study of pulse profiles of short period X-ray pulsars can also be useful for using them as means of interplanetary navigation system. The X-ray pulse profile of this source has been analysed
 for 15 $ \rm years $ (1996--2011). The long term average amplitudes of the first, second and third 
harmonics (and their standard deviation for individual measurements) compared to the fundamental are 36.9 \% (1.7 \%), 13.4 \% (1.9 \%) and 9.4 \% (1.8 \%) 
respectively. Similarly, the phases of the three harmonics (and standard deviations) with respect to the fundamental are  0.36 (0.06), 1.5 (0.2), 2.5 (0.3) 
 $\rm radian $  respectively. We do not find any significant variation of the harmonic components of the pulse profile in comparison to the fundamental.
}

   \authorrunning{P.Pradhan et al. }            
   \titlerunning{X-ray Pulse Profile of PSR B1509-58}  
    \maketitle

\section{Introduction}           
\label{sect:intro}
\noindent
With several sensitive space missions in the last two decades, the detection of rotation-powered pulsars
observable at X-ray energies has increased substantially. 
Pulsars with ages ranging from $10^{3}$ - $7\times10^{9}$ $ \rm years $, magnetic field strength ranging from $10^8$ - $10^{13}$ $ \rm G $ 
and spin periods ranging from 1.6 $ \rm ms $ - 530 $ \rm ms $
have been detected in the X-rays (\citealt{BT97}). Rapid rotation  coupled with strong magnetic fields of 
these pulsars 
results in  rotation-induded electric field  which in turn accelerates the 
$e^+$-$e^-$ pairs in the magnetosphere leading to  high energy ($\gamma$-/X-ray)
 emission either 
by pair-photon cascades initiated by high energy photons above a polar cap (\citealt[]{RS75, DH82, DH96}) 
or in the outer gap (\citealt []{CHR86a, CHR86b, H08, T10, W10}). 
\noindent
Owing to their fast rotation periods, magnetic fields of millisecond and young pulsars at their
 respective light cylinders are comparable even though the surface magnetic fields of the young pulsars
 are nearly five orders of magnitude larger than those of the former. 
 This suggests to us that the magnetic field strength at the light
cylinder may play a role in high energy emission in pulsars (\citealt{W00}).

\noindent
Pulse profiles give us important information about the X-ray emission geometry of the pulsars.  
The average pulse profiles of pulsars, created by folding the light curve of the pulsars with
their respective pulse period, exhibit steady form and are characteristic of every pulsar. 

\noindent
In young pulsars with ages less than 2000 $ \rm years $ like the Crab pulsar, magnetospheric emission
from charged particles accelerated in the neutron star magnetosphere along the curved magnetic field 
lines (Outer Gap model) dominate (\citealt{H08}). The magnetospheric component at X-ray energies is
characterized by strong pulsations, sometimes with several peaked structure (\citealt{BT97}). 
Asymmetric pulse shapes indicate lack of axial symmetry in the emission zone. In millisecond pulsars like PSR 1821-24, 1937+21, 
and J0218+4232, the X-ray emission is dominated by non-thermal processes (\citealt{BT97}). 
Their pulse profiles have narrow peaks and high pulsed fraction.

\noindent
PSR B1509-58 was discovered in X-rays by the Einstein satellite (\citealt{SH82}), and later
 observations in radio (\citealt{MND82}) confirmed
 its 150  $ \rm ms $ period and the highest spin-down rate of $\dot{P}$ $\sim$ 1.5$\times$ $10^{-12}$ 
\rm{s s$^{-1}$} of any known pulsar. 
It has a characteristic age nearly
1700 $ \rm years $, spin-down luminosity of
 $\dot{E}$ = 1.7 $\times$ $10^{37}$ $ \rm {erg ~ s^{-1}}$
 , and dipole magnetic field of 1.5 $\times$ $ 10^{13}$ $ \rm G $. This magnetic field is larger than the magnetar SGR 0418+5729
having the smallest magnetic field less than 7.5 $\times$ $10 ^{12}$ $ \rm G $ (\citealt{Rea10}). Further, its braking index, $n$ is 
less than 3 $ \rm i.e $.,2.839 $\pm$ 0.003 (\citealt{Liv11}). 
This could be due to the magnetic field being non-dipolar (\citealt{KM94}), or a result of pulsar wind that 
carries particles taking angular momentum away from the pulsar causing resultant mass loss (\citealt{MND85}). 
Some other reasons could be having a time variable effective magnetic moment (\citealt{BR88}) or the 
torque function defined by $K$ = - $\dot{\Omega}$/{$\Omega^n$} where $\Omega$ is the spin rate of pulsar, being
time varying (\citealt{AH97}).  

\noindent 
PSR B1509-58 has been observed regularly since 1996 till 2011 with the Proportional Counter Array (PCA)
(\citealt[]{Jah96,Jah06}) on board the Rossi X-ray Timing Explorer (RXTE).
No glitches from this source have been observed so far hence it gives us an opportunity to study the
basic emission properties of this source over a long time without interruption. Also no magnetar like
 X-ray bursts were discovered in this pulsar. 
\noindent
Previous works on this pulsar showed that the X-ray component lags the radio component of the 
pulse by $\sim$ 0.27 period.
The phase relation between the radio and the X-ray pulses, i.e, the arrival time difference
 between the radio and the X-ray pulses have been found to be quite stable (\citealt{L05}).
This lag is energy independent for the range 
2-100 $\rm KeV $ (\citealt{Rot98}) suggesting that the radio and X-ray emission regions are different favouring the 
Outer Gap model. This pulsar is known to emit radiations 
from radio to soft-gamma region suggesting that there may be
new class of pulsars called soft-gamma ray pulsars (\citealt{PP11}).  
\noindent
The timing studies of radio pulsars reveal that they are subjected to a systematic delay or spinning down. 
The measurements of the first and second time derivatives of pulsars’ spin period can 
provide us useful information about the dynamics of rotation of non-accreting neutron stars. 
The timing residuals for PSR B1509-58 from 1996-2010 shows some significant structure.
 Also while the braking index $n$ is stable at long time scales, its variabilty is visible in 
short time scales (\citealt{Liv11}).
This could well be due to changes in the magnetospheric activities of this pulsar which may produce 
corresponding changes in the high energy X-ray emission. 
For PSR B1509-58, no variability in X-ray flux nor any pulse profile
variability has been found when the pulse profiles were compared by a $\chi^{2}$ test, 
while upper limits of 28 \%  was obtained in flux 
change (\citealt{Liv11}). \\ \\
\noindent
In this work we present our analysis of fifteen
years of archival data of PSR B1509-58 observed using RXTE. 
The next section gives the details of the observations and the analysis
techniques used; followed by section 3 where we present the results. 
We then conclude with discussions on the obtained results.

\section{Observation and Data Analysis}
\label{sect:Obs}
\noindent
The Rossi X-ray Timing Explorer (RXTE) was launched on December 30, 1995. 
 It comprised of two pointed instruments, 
the Proportional Counter Array (PCA) in the energy range 2-60 $ \rm KeV $ and the High Energy X-ray 
Timing Experiment (HEXTE) covering the higher energy range 15-250 $ \rm KeV $. In addition, RXTE carried 
an All-Sky Monitor (ASM) that scanned the sky. 
We used data obtained with the Proportional Counter Array (PCA) (\citealt{Jah96}). 
PCA is a collimated array of proportional counters, and is composed of 5 Proportional Counter Units (PCUs) 
with a total photon collection area of 
6500 $ \rm cm^{2}$. Over the years, the average number of detectors available for observation 
decreased and the mission ended in early 2012. 

\begin{figure*}
\begin{center}
 \includegraphics[width=9cm,height=12cm,angle=-90]{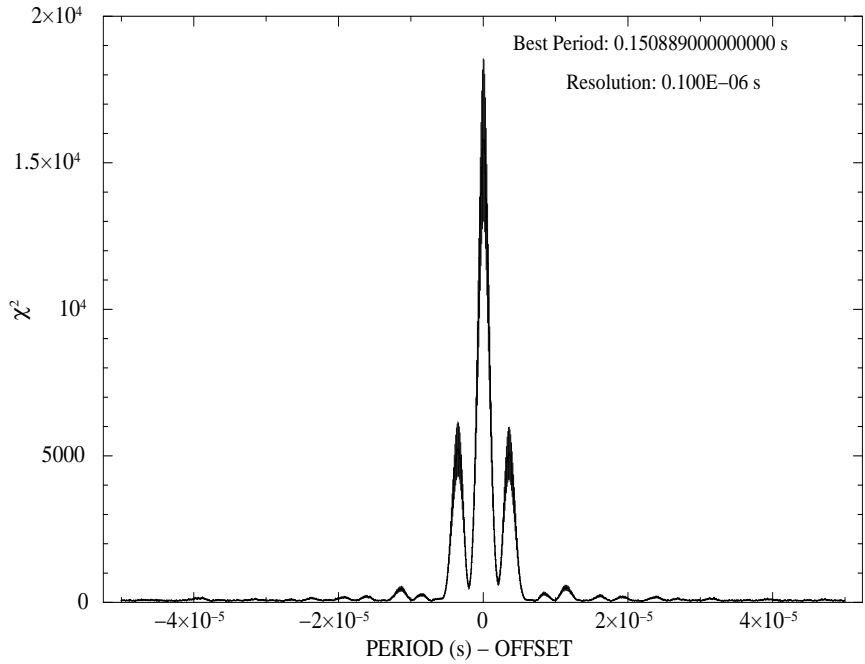}
\end{center}
\caption{The $\chi^{2}$ in the pulse profile for different trial pulse period is plotted here for a sample observation. 
The peak represents the true period.}
\label{efsearch-plot} 
\end{figure*}

\noindent
For PSR B1509-58, archived RXTE-PCA data are available for 15 $ \rm years $ from 1996 to 2011
 which enables us to study its timing characteristics in detail. We used data in Good Xenon 
mode that provides a full timing accuracy of about 1 $ \rm \mu$s. 
From the Science Event files recorded in Good Xenon mode, light curves were created using \texttt{seextrct} 
with a binning of 10 $ \rm ms $. The exposure for the 262 observations used for this analysis were typically of the 
order of kilo seconds.
Barycentric corrections were then made on the light curve so created using \texttt{faxbary}. 
Using the ftools task \texttt{efsearch}, we find the best pulse period for each of the barycenter
corrected light curves.
One such period search is shown in Figure \ref{efsearch-plot}. 
Pulse profiles with 128 phase bins were created by folding the light curves with their 
respective best periods determined with a resolution of $10^{-7}$ $ \rm s $.

\begin{figure*}
\begin{center}
 \includegraphics[width=9cm,height=12cm,angle=-90]{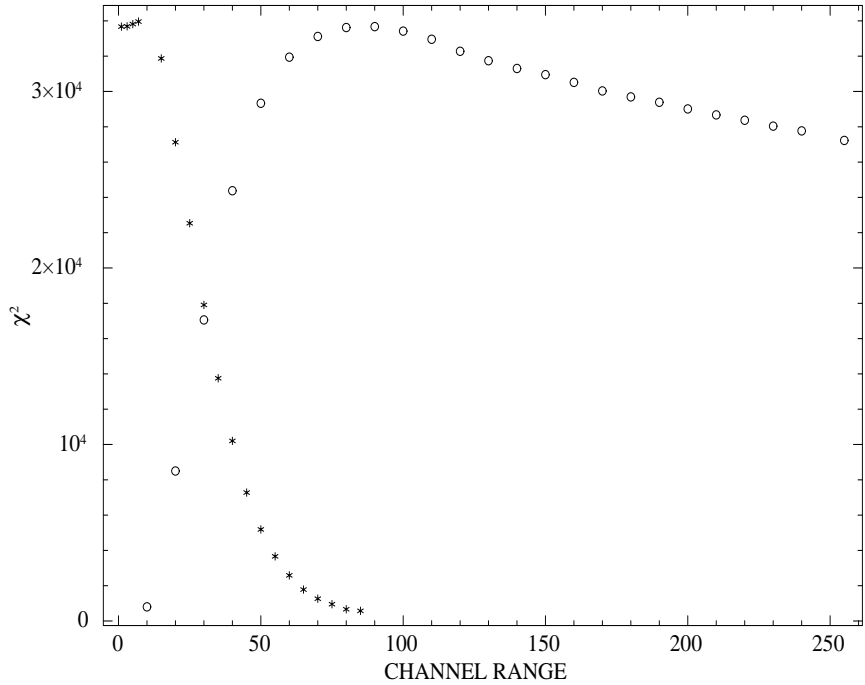}
 \end{center}
\caption {Plot for optimisation of Channel Range. The $\chi^{2}$ of the folded light curve is plotted here for 
different channel ranges.
For the circles, the X-axis represents the highest channel while the lowest channel is set to zero. 
This curve stops rising at about channel no 90.
For the asterisks, the X-axis represents  the lowest channel when the highest channel
is fixed at 90.
A channel range of 3-80 was selected for further analysis which will give highest signal to noise ratio in 
the pulse profile.  }
\label{channel-search} 
\end{figure*}

\begin{figure*}
\begin{center}
 \includegraphics[width=9cm,height=12cm,angle=-90]{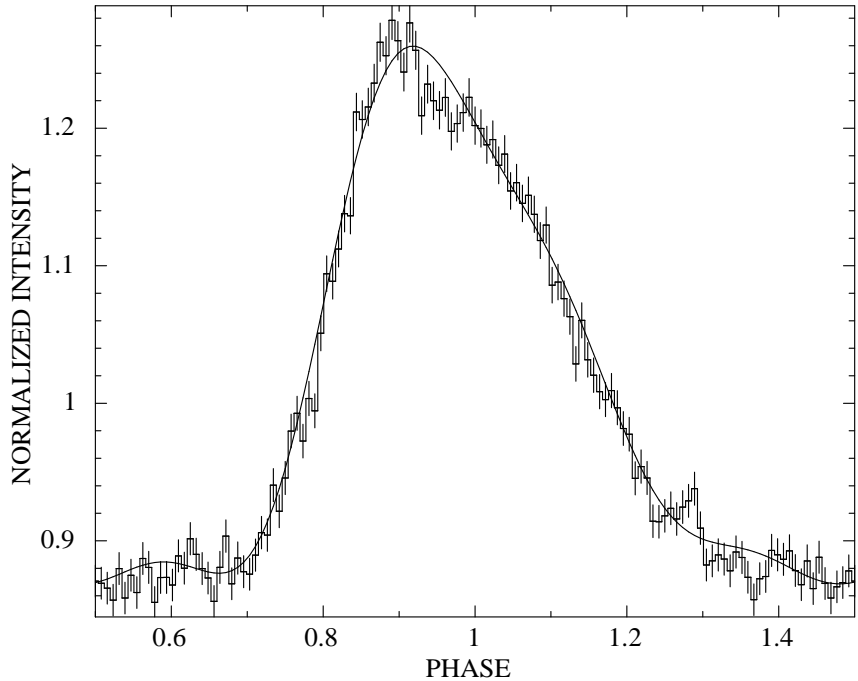}
\end{center}
\caption{A sample pulse profile in the energy range 2-24 $ \rm KeV $. The pulse profile reconstructed by using the fundamental 
and the first three harmonics is overlayed.}
\label{pulse-profile} 
\end{figure*}

\noindent
In order to choose the appropriate energy range, we first searched for channel range with the
maximum signal to noise ratio. The light curves were extracted for different energy intervals for a 
1996 observation and the respective pulse profiles were created. Each profile was fitted with a constant 
and the $\chi^2$ value was noted. 
The greater is the deviation, the larger is 
the value of $\chi^2$, and hence greater is the signal to noise ratio.
This way the channel and hence energy range is optimised for further analysis.
The channel range optimisation plot is given in Figure \ref{channel-search}.
The pulse profiles were created for all the 262 observations taking the energy range to be ~2 to 24 $ \rm KeV $.
A sample pulse profile is given in Figure \ref{pulse-profile}.

\begin{figure*}
\begin{center}
 \includegraphics[width=9cm,height=12cm,angle=-90]{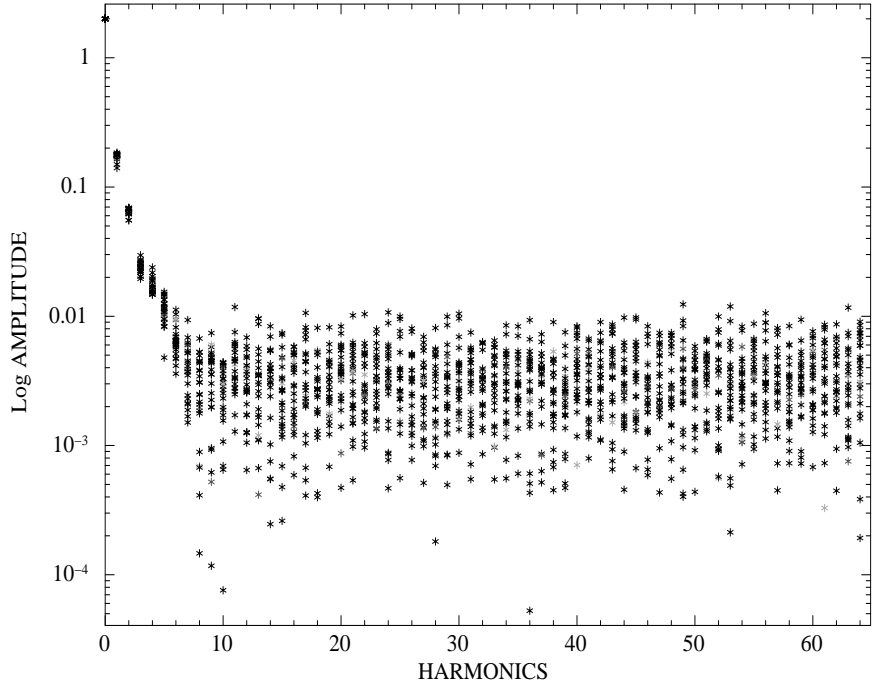}
 \end{center}
\caption{Amplitudes of the harmonics are plotted here for
 25 pulse profiles obtained from observations in 2004.
} 
\label{25-amp}

\end{figure*}

\begin{figure*}
\begin{center}
 \includegraphics[width=9cm,height=12cm,angle=-90]{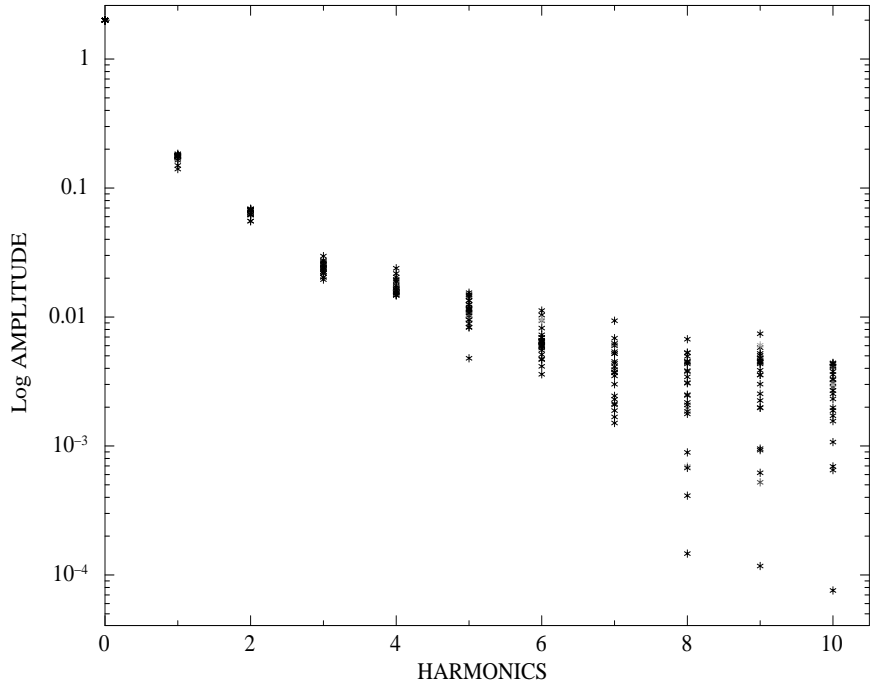}
 \end{center}
\caption{
Same as Figure \ref{25-amp}, only the first 10 harmonics are shown here.}
\label{25-amp1} 
\end{figure*}

\noindent
Next, Fourier decomposition of the pulse profiles were carried out. 
The Fast Fourier Transform (FFT) method will decompose the signal from 
the pulsar it into Fourier components and we get the amplitudes and 
phases of the frequency components the sum of which make the 
individual pulse profiles. It is then possible to calculate the 
relative amplitudes and the phase differences of the harmonics 
with respect to the fundamental for the pulse profile of all the 
years. Should the pulse profile remain unchanged the 
the relative amplitudes and phases of the harmonics with respect to the fundamental will be constant over time and will establish the hypothesis 
that this pulsar has stable X-ray properties on long time scales. \\
\noindent 
Since the pulse profiles have limited statistics, there are definite uncertainties with the measurement of the amplitudes and the phases. 
We have decomposed the individual pulse profiles into 128 Fourier components.
In our analysis, we see that the first four Fourier components have the largest amplitudes 
and therefore have small relative errors and are hence the most significant components.
 We compare only these significant Fourier components. 
This is depicted in Figures \ref{25-amp} \& \ref{25-amp1}.
\noindent
Hence we make a comparison of the relative amplitudes
and phase differences of the first three harmonics with respect to the fundamental.
To make a comparison of phases, first the fundamental is shifted to zero, by subtracting it 
from itself. 
The first harmonic is shifted by twice the fundamental, second harmonic by thrice and the
third harmonic by four times that.
This way the phase differences between 
the first, second and third harmonic and the fundamental is determined. We also calculate the ratio 
of the amplitudes of the first three harmonics with respect 
to the fundamental.  \\

\section{Results}
\label{sect:results}
\begin{figure*}
\begin{center}
 \includegraphics[width=9cm,height=12cm,angle=-90]{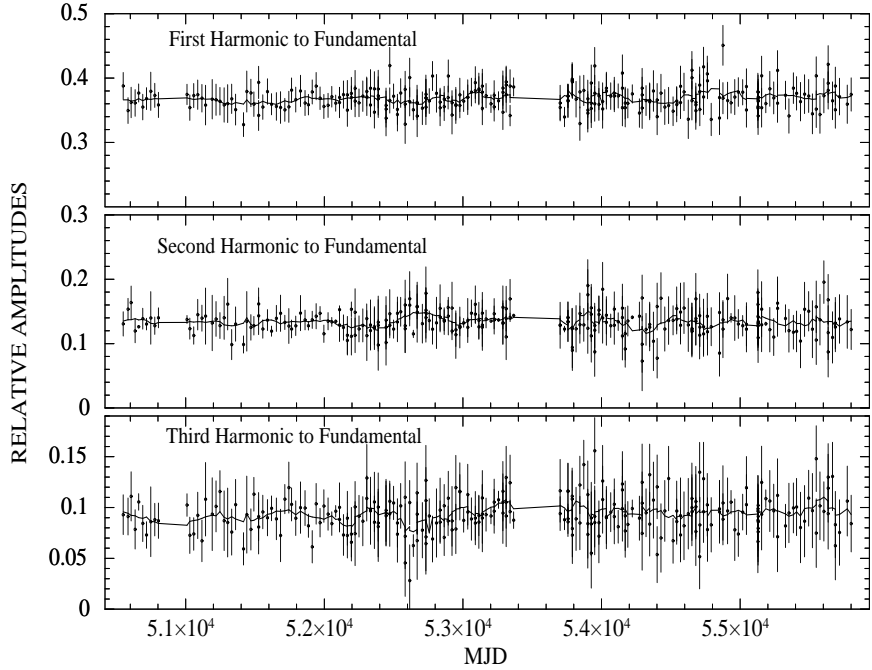}
 \end{center}
\caption{The line represents the running mean over nine points of the relative amplitudes of the first three harmonics with respect to the fundamental while 
the points represent the actual values. The X-axis represents the MJD. The error bars were determined using the total number of photons in each pulse profile, results
from a Monte Carlo simulation of the profiles and a scaling factor described in section \ref{sect:results}.}
\label{ra}
\end{figure*}

\begin{figure*}
\begin{center}
 \includegraphics[width=9cm,height=12cm,angle=-90]{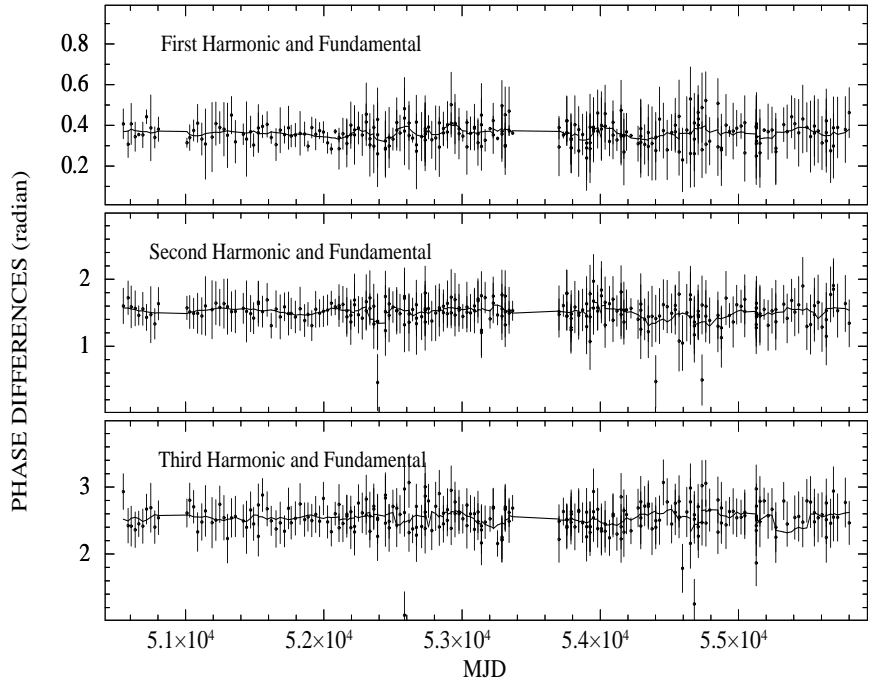}
 \end{center}
\caption{The line represents the running mean over nine points of the phase difference of the first three harmonics and the fundamental while the points 
represent the actual values. The X-axis represents the MJD. The error bars are calculated in the same way as Figure \ref{ra}.} 
\label{pd}
\end{figure*}

\noindent
The relative amplitudes of the first,second and third harmonics and their fundamental for all the 
observations are shown in Figure \ref{ra} and the phase differences of the first three harmonics with respect to the fundamental in Figure \ref{pd}. 
The long term average amplitudes of the first, second and third 
harmonics (and their standard deviation for individual measurements) compared to the fundamental are 36.9 \% (1.7 \%), 13.4 \% (1.9 \%) and 9.4 \% (1.8 \%) 
respectively. Similarly, the phases of the three harmonics (and standard deviations) with respect to the fundamental are  0.36 (0.06), 1.5 (0.2), 2.5 (0.3) $\rm radian $ 
respectively.\\
To calculate the errors on the measured amplitudes and phases of the harmonics we have carried out a Monte Carlo simulation. Pulse profiles were simulated with an 
intrinsic profile same as that shown with the solid line in Figure \ref{pulse-profile} but with a Gaussian deviation in each bin due to limited photon statistics. The same parameters were then measured from 
the simulated pulse profiles. The process was carried out 10,000 times and the standard deviation of each of the parameters
obtained from the 10,000 simulations was taken to be the error on that parameter. The error
obtained in a given pulse profile however depends on the total number of photons used to create
the pulse profile. We have therefore estimated the dependence of the errors on the total number of
photons used to create the pulse profiles by repeating the process for different number of total
photons. As expected, a power-law dependence with index -1/2 was found between the errors and
the total number of photons. The total number of X-ray photons, which is known for each of the
observations was then used along with the above results to determine the error of the parameters
for each of the observations. It was however found that there is a larger scatter in the measured
parameters around the mean value compared to those obtained from the Monte Carlo simulation.
We have measured the standard deviation of the parameters obtained from the 25 observations
carried out in 2004 and obtained a scaling between the scatter and the errors estimated from the
Monte Carlo simulation. All the error bars were multiplied by this scaling factor and are shown
in Figures \ref{ra} and \ref{pd}.\\
To investigate if there is an underlying trend, we have 
overplotted a nine point running average for each of the plots. The running mean indicates that on a longer timescale of about a year, there could be 
some systematic variation in the pulse profiles with upper limit on  percentage variation of the relative
amplitudes of the first three harmonics to the fundamental being within 4 \%, 13 \% and 17 \% and phases differences of first three harmonics from the
fundamental being within 0.05, 0.15 and 0.14 $ \rm radian $ of the respective
long term averages. From Figures \ref{ra} and \ref{pd}, it is also evident that there is a greater variation of phase difference and relative amplitudes for the later years data
of RXTE from the mean. This is probably due to the reduced sensitivity of the PCA resulting from loss of PCUs in the subsequent years of operation.

\section{Discussion}
\label{sect:discussion}
\noindent
Pulsars show many types of flux variation on different time scales. The short term flux variation being
bursts (\citealt{LVT93}) and glitches. (\citealt{LPS95}).  
\noindent
No glitches has been observed for PSR B1509-58 while the younger Crab pulsar
had 24 glitches in 42 $ \rm years $ (\citealt{ELSK11}) implying that the former has a higher internal 
temperature than the latter (\citealt{ML90}). 
Unlike the Crab and the Vela pulsar, PSR B1509-58 exhibits only an asymmetric, 
broad X-ray pulse suggesting that X-ray pulse emission region is closer to neutron star for 
PSR B1509-58 than for the other two (\citealt{K91}).
The pulse profile analysis of PSR B1509-58 done here for the observations during the years 1996--2011 
indicate some scatter in the amplitudes and phases. This is however
not detectable by eye on pulse profiles. 

\noindent
It is worthwhile mentioning that the change in the radio pulse behaviour of pulsars during time as
short as their periods is considered a characteristic of radio emission solely.
For the rotation powered pulsars, it is not possible to compare single X-ray pulses due to limited
photon statistics. However, connection between the spin-down characteristics and the radio emission 
that is well known, can also be probed for the high energy emission. In this context, it is interesting 
to note the behaviour of PSR B1931+24. This intermittent pulsar stopped emitting for 
days during which it spins down half rapidly (\citealt{Y13}). 
For PSR J1841-0500 (\citealt{C12}) and PSR J1832+0029 (\citealt{L07})
too, changes in spin down rate is associated with the variations in their average radio profiles.  
For PSR B1509-58, we are investigating if changes in the spin down 
characteristics are associated with any changes in the high energy emission properties.

\noindent
We have made a quantitative estimate of any possible changes in 
the pulse profile by carrying out a Fourier decomposition and put upper limits of 
36.9 \%, 13.4 \%, 9.4 \% on the amplitudes and 0.36, 1.5, 2.5 $\rm radian $ on the phase 
of the first three harmonics with respect to the fundamental. 
A similar study of the pulse profile of the Crab pulsar, but using different analytical expression 
for the profile, showed no pulse profile variation over a decade (\citealt{CB11}). 

\noindent
The pulse profile stability of short period X-ray pulsars has 
an interesting application in interplanetary spacecraft navigation. By comparing the delay between the pulse arrival time
measured onboard the spacecraft and the predicted arrival times at an inertial reference like the 
solar system barycentre, we get the relative position of the pulsar
along the line of sight towards the pulsar. Three-dimensional position information of the spacecraft can
be  obtained likewise from the same information of at least
three different pulsars (\citealt{B10}). 
This way, if we have a short period X-ray pulsar with a stable pulse profile, 
they can be used for navigation of spacecrafts. Also, decomposing the whole pulse profile into its 
fourier components as discussed in this work has an advantage over using only the arrival time of the
pulse peak (\citealt{S06}). \\
Compared to the millisecond pulsars, PSR B1509-58 is much slower and will not give high position resolution for navigation. However,
to be useful for the purpose of navigation, it is necessary for the X-ray pulse profile of rotation powered pulsars to be stable.
After the Crab pulsar, PSR B1509-58 is the only bright source with many X-ray observation that can be used to investigate the pulse profile
stability. In the present work, we have given upper limits to its profile variation in terms of stability of the harmonic components with respect 
to the fundamental.\\

\normalem
\begin{acknowledgements}
The data for this work has been obtained through the High Energy Astrophysics Science Archive (HEASARC) 
Online Service provided by NASA/GSFC.
Also thanks to the hospitality provided by Raman Research Institute to PP and PP 
deeply thanks Chandreyee Maitra for her support in this analysis.
 \end{acknowledgements}

\bibliographystyle{raa}
\bibliography{ms1673}

\label{lastpage}
\end{document}